\documentclass[11pt]{article}
\pdfoutput=1 
\usepackage{amssymb}
\usepackage{amsmath}
\usepackage{amstext}
\usepackage{graphicx,epsfig}
\usepackage{epsfig}
\usepackage{verbatim} 
\usepackage{fancybox}
\usepackage{color}
\usepackage{ulem}
\usepackage{enumitem}
\usepackage{youngtab}
\usepackage{subfigure}
\usepackage{bbm}
\usepackage{parskip}
\usepackage[numbers,sort&compress]{natbib}
\usepackage{ytableau}
\usepackage[utf8]{inputenc}

\usepackage{tikz}
\usetikzlibrary{positioning, trees,decorations, decorations.markings, decorations.pathmorphing, arrows, shapes.geometric, snakes, patterns}
\pgfdeclaredecoration{complete sines}{initial}
{
\state{initial}[
width=+0pt,
next state=upsine,
persistent precomputation={\pgfmathsetmacro\matchinglength{
\pgfdecoratedinputsegmentlength / int(\pgfdecoratedinputsegmentlength/\pgfdecorationsegmentlength)}
\setlength{\pgfdecorationsegmentlength}{\matchinglength pt}
}] {}
\state{upsine}[width=\pgfdecorationsegmentlength,next state=downsine]{
\pgfpathsine{\pgfpoint{0.25\pgfdecorationsegmentlength}{0.5\pgfdecorationsegmentamplitude}}
\pgfpathcosine{\pgfpoint{0.25\pgfdecorationsegmentlength}{-0.5\pgfdecorationsegmentamplitude}}
}
\state{downsine}[width=\pgfdecorationsegmentlength,next state=upsine]{
\pgfpathsine{\pgfpoint{0.25\pgfdecorationsegmentlength}{-0.5\pgfdecorationsegmentamplitude}}
\pgfpathcosine{\pgfpoint{0.25\pgfdecorationsegmentlength}{0.5\pgfdecorationsegmentamplitude}}
}
\state{final}{}
}

\newcommand{\Comment}[1]{{}}
\definecolor{darkblue}{rgb}{0.15,0.35,0.55}
\definecolor{reddish}{rgb}{0.65, 0.2, 0.2}
\usepackage[linktocpage=true]{hyperref}
\hypersetup{
colorlinks=true,
citecolor=darkblue,
linkcolor=reddish,
urlcolor=darkblue,
pdfauthor={},
pdftitle={},
pdfsubject={}
}

\definecolor{green3}{RGB}{44, 160, 44}

\newcommand{\be}{\begin{equation}}
\newcommand{\ee}{\end{equation}}
\newcommand{\bea}{\begin{eqnarray}}
\newcommand{\eea}{\end{eqnarray}}
\newcommand{\beas}{\begin{eqnarray*}}
\newcommand{\eeas}{\end{eqnarray*}}

\definecolor{darkred}{rgb}{0.7,0.3,0.3}
\definecolor{darkgreen}{rgb}{0.2,0.7,0.3}
\definecolor{lightgreen}{rgb}{.816,.94,.753}
\definecolor{greyish}{rgb}{.8,.8,.8}
\definecolor{darkblue2}{rgb}{0.3,0.4,0.9}
\usepackage{pifont}
\usepackage{xcolor,colortbl}

\def\({\left(}
\def\){\right)}

\newcommand{\la}{\langle}
\newcommand{\ra}{\rangle}

\def\gsim{ \lower .75ex \hbox{$\sim$} \llap{\raise .27ex \hbox{$>$}} }
\def\lsim{ \lower .75ex \hbox{$\sim$} \llap{\raise .27ex \hbox{$<$}} }

\def\xyma{\xymatrix@M.7em}
\def\xymas{\xymatrix@M.1em}

\title{}
\author{}

\numberwithin{equation}{section}

\begin{document}
\tikzset{
photon/.style={decorate, decoration={snake}, draw=magenta},
graviton/.style={decorate, decoration={snake}, draw=black},
sgal/.style={decorate , dashed, draw=black},
scalar/.style={decorate , draw=black},
mgraviton/.style={decorate, draw=black,
decoration={coil,amplitude=4.5pt, segment length=7pt}}
electron/.style={draw=blue, postaction={decorate},
decoration={markings,mark=at position .55 with {\arrow[draw=blue]{>}}}},
gluon/.style={decorate, draw=magenta,
decoration={coil,amplitude=4pt, segment length=5pt}} 
}
\renewcommand{\thefootnote}{\fnsymbol{footnote}}
~

\begin{center}
{\Large \bf Scale vs. Conformal Invariance at the \\ IR Fixed Point of Quantum Gravity\\
}
\end{center} 

\vspace{1truecm}
\thispagestyle{empty}
\centerline{\Large Kara Farnsworth,${}^{\rm a,}$\footnote{\href{mailto:kmfarnsworth@gmail.com}{\texttt{kmfarnsworth@gmail.com}}} Kurt Hinterbichler,${}^{\rm a,}$\footnote{\href{mailto:kurt.hinterbichler@case.edu} {\texttt{kurt.hinterbichler@case.edu}}} Ond\v{r}ej Hul\'{i}k,${}^{\rm a,b,}$\footnote{\href{mailto:ondra.hulik@gmail.com} {\texttt{ondra.hulik@gmail.com}}} }

\vspace{.5cm}

\centerline{{\it ${}^{\rm a}$CERCA, Department of Physics,}}
\centerline{{\it Case Western Reserve University, 10900 Euclid Ave, Cleveland, OH 44106}} 
\vspace{.25cm}

\centerline{{\it ${}^{\rm b}$ Theoretische Natuurkunde, Vrije Universiteit Brussel,}}
\centerline{{\it Pleinlaan 2, B-1050 Brussels, Belgium}} 

\vspace{1cm}
\begin{abstract}
\noindent

We examine the question of scale vs. conformal invariance for the linearized Einstein-Hilbert action, which describes the IR fixed point of quantum gravity. In $D = 4$, although the action is not conformally invariant in the usual sense, we explicitly show that the theory is a conformal field theory at the level of correlation functions. In higher dimensions, we show that the theory is scale but not conformally invariant, but can be embedded into a larger non-unitary conformal field theory, analogous to what has been found for Maxwell theory in $D>4$. We give evidence that similar statements are true for all free higher spin theories.

\end{abstract}


\setcounter{tocdepth}{2}
\renewcommand*{\thefootnote}{\arabic{footnote}}
\setcounter{footnote}{0}

\newpage

\section{Introduction}

In the infrared (IR) limit, we expect any quantum field theory to approach a fixed point of the renormalization group (RG) flow. The fixed point is by definition scale invariant, however we often find that this symmetry is enhanced to full conformal invariance.  This symmetry enhancement has been proven for unitary theories with a stress-energy tensor in spacetime dimension $D = 2$ \cite{Polchinski:1987dy} and there is strong evidence it is also true in $D = 4$ \cite{Dymarsky:2013pqa,Dymarsky:2014zja,Yonekura:2014tha}, with the exception of very particular free theories.
 (For a review of the topic of scale vs. conformal invariance see \cite{Nakayama:2013is}.)

Quantum gravity is not expected to conform to the standard framework of quantum field theory; for example it is not expected to contain any true gauge-invariant local operators. Nevertheless, other observables, such as the $S$-matrix on flat space, are expected to be well-defined at all scales, and one can still ask about an RG flow relating these observables at different energy scales. The mysteries of quantum gravity are typically associated with the ultraviolet (UV) of this flow, but we do know something about the IR limit.  In a low energy expansion, the S-matrix organizes as an expansion in positive powers of the external energies $E$, over the Planck mass $M_P$ and possibly other masses $m$ associated with massive particles that have been integrated out.  The lowest powers of $E$ come from tree level diagrams with vertices drawn from the Einstein-Hilbert action, the next lowest powers from 1-loop diagrams with vertices drawn from Einstein-Hilbert and trees with one vertex from 4-derivative terms, and so on (see the power counting rules in \cite{Burgess:2007pt}).  Thus $\sim E/ M_P,\ E/m$ are the effective couplings, and as we take $E$ to zero to approach the infrared fixed point, the couplings go to zero.  Thus the IR fixed point theory
is the free linearized Einstein-Hilbert action.

The linearized Einstein-Hilbert action is scale invariant in any $D$, as required for a fixed point theory, but it does not manifest full conformal symmetry (except in $D\leq 2$ where the action is trivial).  For $D\geq 4$, this theory possesses non-trivial gauge-invariant local operators with which we can investigate this symmetry: the linearized Weyl tensor and its derivatives. In this paper we study the conformal invariance of the linearized Einstein-Hilbert action by explicitly computing correlation functions of the Weyl tensor operator.

In $D=4$, there is evidence to suggest the theory possesses a hidden conformal invariance. For example, the massless wave equation for higher spin fields is conformally invariant in $D = 4$ \cite{Bracken:1982ny,Eastwood:1987ki,Siegel:1988gd,Erdmenger:1997gy,Shaynkman:2004vu,Nakayama:2013is,Barnich:2015tma,Flores:2017yyj,Flores:2017ubo}, and so correlation functions formulated from a generalized free field prescription in this case are consistent with those of a conformal primary operator.  However we are unaware of an explicit derivation of the conformal correlation functions of these operators directly from the action. We will present such a derivation, showing that the correlation functions are conformally invariant, and thus demonstrating that the IR fixed point of quantum gravity in $D=4$ is a conformal field theory (CFT).  This CFT does not possess a stress tensor, or any relevant or marginal local scalar operators.

For $D>4$, we show from the correlation functions that the theory is scale but not conformally invariant. Furthermore, we find that we can embed it into a larger non-unitary conformal theory by fixing a `conformal gauge' in the action. 

These statements about the linearized Einstein-Hilbert action, which describes a massless spin-2 particle, are analogous to what happens in Maxwell theory \cite{El-Showk:2011xbs}, which describes a massless spin-1 particle.  The difference is that the conformal symmetry of the Maxwell theory in $D=4$ is visible in standard form directly in the action.  We argue in Section \ref{HSsection} that similar statements are true for all the free higher spin fields $s>2$: they are conformal in $D=4$ (though the action does not have standard conformal symmetry), and in $D>4$ they are scale invariant but not conformally invariant but should be embeddable into larger non-unitary CFTs.  

\textbf{Conventions:} $D$ is the spacetime dimension, and we use the mostly plus metric signature. Indices are (anti)symmetrized with weight one.

\section{The IR fixed point of quantum gravity}

Quantum gravity, coupled to any other massive states, is described at energies much below those of the massive states by the Einstein-Hilbert effective field theory
\be S= {M^{D-2}_P\over 2}\int d^Dx \sqrt{-g} R+{\cal O}(\partial^4).\label{einsteinhilbertlee}\ee
In the deep IR, the Planck mass $M_P$, and any other mass scales, flow to infinity and we reach an IR fixed point, described by the free {Fierz-Pauli action} \cite{Fierz:1939ix},
\begin{align}
\label{masslessefreeaction} 
S=&\ \int d^D x \left[-\tfrac{1}{2}\partial_\lambda h_{\mu\nu}\partial^\lambda h^{\mu\nu}+\partial_\mu h_{\nu\lambda}\partial^\nu h^{\mu\lambda}-\partial_\mu h^{\mu\nu}\partial_\nu h+\tfrac{1}{2}\partial_\lambda h\partial^\lambda h\right].
\end{align}
Here $h_{\mu\nu}$ is the canonically normalized fluctuation of the graviton around flat space, $h_{\mu\nu}=\frac{1}{2}M_P^{{1\over 2}(D-2)}(g_{\mu\nu}-\eta_{\mu\nu})$. The action \eqref{masslessefreeaction} has the gauge symmetry
\be\label{gaugesymorig} \delta h_{\mu\nu}=\partial_{\mu}\xi_\nu+\partial_{\nu}\xi_\mu,\ee
with a vector gauge parameter $\xi_\mu$,
which is the linearization of the diffeomorphism invariance of \eqref{einsteinhilbertlee}. In $D\geq 4$ this action describes a single propagating massless spin-2 particle. In $D=3$ it is purely constraint and has no propagating degrees of freedom and in $D\leq 2$ it is a total derivative. 

The action \eqref{masslessefreeaction} is manifestly invariant under a standard scale transformation, 
\be \delta h_{\mu\nu}= -\left(x^\lambda \partial_\lambda+\Delta \right) h_{\mu\nu} ,
\ee
with scaling weight
\be \Delta =\tfrac{1}{2}(D-2),\ee
inherited from the dimension of $M_P$.  Scale invariance such as this is necessary at a fixed point of an RG flow.

Scale invariance at a fixed point is often enhanced to conformal invariance, which adds the special conformal transformations to the symmetry algebra. The standard form of a special conformal transformation on a spin-2 primary field is
\begin{align}
\label{specialconformalhe} 
\delta^\sigma h_{\mu\nu}=&\ \left( -2x^\sigma x^\lambda\partial_\lambda+x^2 \partial^\sigma -2 x^\sigma\Delta\right)h_{\mu\nu}-2x_\lambda ({\cal J}^{\sigma\lambda})_{\mu\nu}^{\ \ \alpha\beta}h_{\alpha\beta} ,
\end{align}
where $({\cal J}^{\sigma\lambda} )_{\mu\nu}^{\ \ \alpha\beta}= ( 2\delta^{[\sigma}_\mu \eta^{\lambda]\alpha}) \delta_\nu^\beta+ \delta_\mu^\alpha( 2\delta^{[\sigma}_\nu \eta^{\lambda]\beta})$ is the Lorentz generator in the tensor representation. 
The action \eqref{masslessefreeaction} fails to be invariant under \eqref{specialconformalhe} (except in $D\leq 2$ where the action is a trivial total derivative), so it would seem that the IR fixed point of quantum gravity is an example of a scale but not conformally invariant theory\footnote{Note that this is generally not considered a {\it real} counterexample to the claim that scale invariance implies conformal invariance, since, as we will see below, the theory does not admit a stress-energy tensor \cite{Dorigoni:2009ra}. Nevertheless, we can still ask about the global conformal invariance of correlation functions of other local operators, the approach we will take here.}.

\section{Scale vs. conformal invariance of correlation functions} \label{scco}

The full theory of quantum gravity is not generally expected to have any true gauge-invariant local operators. Nevertheless, the linearized fixed-point theory \eqref{masslessefreeaction} does possess gauge-invariant local operators, namely the linearized Weyl tensor,
\be W_{\mu_1\mu_2\mu_3\mu_4}= 
-4\, {\cal P}_{\mu_1\mu_2\mu_3\mu_4}^{\nu_1\nu_2\nu_3\nu_4}\partial_{\nu_1}\partial_{\nu_3} h_{\nu_2\nu_4}.\label{linearweylopeee}\ee
The tensor ${\cal P}_{\mu_1\mu_2\mu_3\mu_4}^{\nu_1\nu_2\nu_3\nu_4}$ is the projection tensor onto the space of tensors with the algebraic symmetries of the Weyl tensor (anti-symmetric in $\mu_1\mu_2$ and in $\mu_3\mu_4$, 
vanishes when anti-symmetrized among $\mu_1\mu_2\mu_3$, and fully traceless).\footnote{The projector can be constructed by writing an ansatz consisting of a linear combination of terms with four $\eta$'s, with all possible index placements, demanding the $\mu$'s and $\nu$'s satisfy the symmetries of the Weyl tensor, demanding symmetry under exchanging the $\mu$'s with the $\nu$'s, and demanding that the operator squares to itself. This uniquely fixes the ansatz and gives the projector.} The Weyl operator \eqref{linearweylopeee} has dimension ${1\over 2}(D+2)$ (the linearization of the usual fully non-linear dimension $2$ Weyl tensor is given by ${M_P^{-{1 \over 2}(D-2)}} W_{\mu_1\mu_2\mu_3\mu_4}$).

The linear Weyl operator \eqref{linearweylopeee} is invariant under the gauge transformations \eqref{gaugesymorig}, as well as under linearized Weyl transformations
\be \delta h_{\mu\nu}=\Lambda \eta_{\mu\nu},\label{weylsymmlinee}\ee
with scalar gauge parameter $\Lambda$. 
Note that the linearized Ricci tensor vanishes on the equations of motion, so its correlators will not have any support at separated points, making the linearized Weyl tensor the only basic gauge-invariant local operator. 
All other gauge-invariant local operators are constructed from derivatives and products of $W_{\mu_1\mu_2\mu_3\mu_4}$, modulo equations of motion.

In $D=3$, the Ricci tensor is the only gauge-invariant possibility (the Weyl tensor vanishes identically, and the Cotton tensor is written in terms of the Ricci tensor). Since the Ricci tensor vanishes on-shell, there are no gauge-invariant local operators in $D=3$, and the theory is topological. Since the theory is a trivial total derivative in $D\leq 2$, we will thus restrict to $D\geq 4$. 

To compute correlation functions we must first gauge-fix the action. Following the Faddeev-Popov procedure and ignoring the decoupled ghosts, we add a general gauge-fixing term which depends on two gauge-fixing parameters, $\xi_1$ and $ \xi_2$,
\begin{align}
\label{laggaugefixxie} S_{\rm g.f.}=&\ \int d^D x \bigg[-\tfrac{1}{2}\partial_\lambda h_{\mu\nu}\partial^\lambda h^{\mu\nu}+\partial_\mu h_{\nu\lambda}\partial^\nu h^{\mu\lambda}-\partial_\mu h^{\mu\nu}\partial_\nu h+\tfrac{1}{2}\partial_\lambda h\partial^\lambda h-\tfrac{1}{\xi_1}\left(\partial^\nu h_{\mu\nu}-\tfrac{1}{2}\xi_2 \partial_\mu h\right)^2\bigg].
\end{align}
The choice $\xi_1=\xi_2=1$ is the usual Lorentz/de Donder gauge. The only value which is not allowed is $\xi_2=2$, which fails to fix the gauge: for this value there is a residual gauge transformation $\delta h_{\mu\nu}=\partial_\mu\partial_\nu\chi$ with a scalar gauge parameter $\chi$, which is enough to render the propagator non-invertible.

\subsection{$D>4$}
\label{d>4}

When $D>4$, a useful choice for the gauge-fixing parameters is
\be \xi_1={D+2\over D-2},\ \ \xi_2={D+4\over D}. \label{conformalgaugechoice}\ee
After making this choice, we decompose $h_{\mu\nu}$ into a traceless and trace part,
\be h_{\mu\nu}=\tilde h_{\mu\nu}+\phi\, \eta_{\mu\nu},\ \ \ \tilde h^\mu_{\ \mu}=0,\ee
and the action  \eqref{laggaugefixxie} becomes
\begin{align} \label{laggaugefixxie2} S_{\rm g.f.}=&\ \int d^D x \bigg[-\tfrac{1}{2}\partial_\lambda \tilde h_{\mu\nu}\partial^\lambda \tilde h^{\mu\nu}+\tfrac{4}{D+2} \partial_\mu \tilde h_{\nu\lambda}\partial^\nu \tilde h^{\mu\lambda}+\tfrac{(D-4)(D-2)}{4}(\partial\phi)^2\bigg].
\end{align}

This action is now conformally invariant \cite{Anselmi:1999bu}, with both $\tilde h_{\mu\nu}$ and $\phi$ transforming as primary fields of weight $\Delta =\tfrac{1}{2}(D-2)$. The two-point functions take the standard form fixed by conformal symmetry:
\bea 
\label{hconf}
\la \phi(x)\phi(0)\ra= \frac{c_\phi }{ x^{2\Delta}},\ \ 
\la \tilde h_{\mu_1\mu_2}(x) \tilde h^{\nu_1\nu_2}(0)\ra= \frac{c_h}{x^{2\Delta}}{\cal I}_{\mu_1\mu_2}^{\nu_1\nu_2},\ \ \ 
\la \tilde h_{\mu_1\mu_2}(x)\phi(0)\ra=0,\ \ \ 
\eea
where
\be
c_\phi =-\frac{\Gamma\left(\tfrac{D}{2}-1\right)}{2(D-4)(D-2)\pi^{D/2}},\quad{} c_h = \frac{D\,\Gamma\left(\tfrac{D}{2}-1\right)}{4(D-4)\pi^{D/2}},\ \ \ \Delta =\tfrac{1}{2}(D-2),
\ee
and 
\begin{align}
{\cal I}_{\mu_1\mu_2}^{\nu_1\nu_2} \equiv \tfrac{1}{2} ( I_{\mu_1}^{ \nu_1} I_{\mu_2}^{ \nu_2}+I_{\mu_1}^{ \nu_2} I_{\mu_2}^{ \nu_1}) - \tfrac{1}{D} \eta_{\mu_1\mu_2} \eta^{\nu_1\nu_2}\, ,\ \ \ 
I^{\mu\nu}&\equiv \eta^{\mu\nu}-2{x^\mu x^\nu\over x^2}\,, \label{Imunudefe}
\end{align}
is the standard spin-2 tensor structure required by conformal invariance. The choice \eqref{conformalgaugechoice} is the only choice for which the gauge-fixed action is conformally invariant in the standard way.

From this we can compute the desired gauge-invariant correlator of Weyl operators,
\begin{align}
\label{WWcorrelatoree}
\la W_{\mu_1\mu_2\mu_3\mu_4}(x) W_{\nu_1\nu_2\nu_3\nu_4}(0)\ra= 16\, {\cal P}_{\mu_1\mu_2\mu_3\mu_4}^{ \rho_1\rho_2\rho_3\rho_4 } {\cal P}_{\nu_1\nu_2\nu_3\nu_4}^{ \sigma_1\sigma_2\sigma_3\sigma_4 } \partial_{\rho_1} \partial_{\rho_3} \partial_{\sigma_1} \partial_{\sigma_3} \la \tilde h_{\rho_2\rho_4}(x) \tilde h_{\sigma_2\sigma_4}(0)\ra.
\end{align} 
Though we computed this using the gauge choice \eqref{conformalgaugechoice}, the answer is the same in any gauge because the Weyl operators are gauge invariant.
Note that the trace field $\phi$ does not contribute due to the invariance \eqref{weylsymmlinee} of the Weyl tensor. 

The correlator \eqref{WWcorrelatoree} satisfies two conservation conditions: it is both conserved and dual conserved,
\bea 
\partial^{\mu_1} \la W_{\mu_1\mu_2\mu_3\mu_4}(x) W_{\nu_1\nu_2\nu_3\nu_4}(0)\ra&=&0, \label{conservqtioncondiee1} \\
\partial_{[\mu_5 }\la W_{\mu_1\mu_2]\mu_3\mu_4}(x) W_{\nu_1\nu_2\nu_3\nu_4}(0)\ra&=&0. \label{conservqtioncondiee2}
\eea
In fact, starting with a general scale invariant ansatz for a two-point function with the correct index symmetries and scaling dimension of the Weyl operator, the first conservation condition \eqref{conservqtioncondiee1} fixes the structure to be that of \eqref{WWcorrelatoree}, after which the second condition \eqref{conservqtioncondiee2} is automatically satisfied. 

The correlator \eqref{WWcorrelatoree} is not of the conformally invariant form for a conformal primary with the Lorentz properties of the Weyl tensor, and instead has the form of a descendent of a primary spin-2 tensor \eqref{hconf}, as manifest by \eqref{WWcorrelatoree}. However, $W$ cannot be a true descendent in the gauge-invariant theory, since it is the lowest dimension operator in the theory (anything lower is not gauge-invariant). Therefore the gauge-invariant sector of the theory is not conformally invariant, only scale invariant.

The equation of motion of $h_{\mu \nu}$ fails to be conformally invariant, however the choice of conformal gauge makes the equation of motion for $\tilde{h}_{\mu\nu}$ conformally invariant.  This in turn allows us to embed the theory in a larger conformal theory, simply by adding in the gauge-fixed $\tilde h_{\mu\nu}$ as a conformal primary, making $W$ a true descendent operator.  However this larger theory is not unitary, because the dimension of $\tilde h_{\mu\nu}$ is ${1\over 2}(D-2)$, which violates the unitarity bound $\Delta\geq D$ for a spin-2 primary \cite{Minwalla:1997ka}. Note that the conservation conditions \eqref{conservqtioncondiee1}, \eqref{conservqtioncondiee2} are level-3 shortening conditions for the conformal Verma module descended from the primary $\tilde h_{\mu\nu}$.

The gauge-invariant, scale invariant theory on its own is unitary. Since it is free, all the correlators of all gauge-invariant local operators can be computed by Wick contraction from the basic correlator \eqref{WWcorrelatoree}. Note that this theory does not have a stress tensor: the stress tensor would have to be a gauge-invariant rank-2 tensor operator of dimension $D$. Such an operator cannot be formed from derivatives of $W$: because of the conservation condition \eqref{conservqtioncondiee1} and tracelessness of $W$, there is no rank-2 tensor that can be formed (the only representation that can be formed with $d$ derivatives is a traceless Young tableaux with one row of length $d+2$ and another of length 2, as manifested in the unfolded formulation of the field equations \cite{Bekaert:2004qos}). Thus the stress tensor would have to involve at least 2 powers of $W$, but this would have dimension at least $D+2$, which is too large to be a stress tensor.  The absence of a stress tensor in this theory also follows from the Weinberg-Witten theorem \cite{Weinberg:1980kq}, which forbids a theory containing both a massless spin-2 particle and a stress tensor. Quantum gravity, which has a massless spin-2 particle in the IR, cannot be a standard field theory with a local stress tensor. 

We also see that there is no scalar primary operator with dimension $\Delta \leq D$ in the gauge-invariant theory, meaning there is no relevant operator that can be used to define an RG flow away from this theory. Therefore this theory cannot be the UV fixed point of a standard quantum field theory, with an RG flow initiated by deforming the theory with relevant local operators. 

\subsection{$D=4$}
\label{d=4}

In $D=4$ the conformal gauge choice \eqref{conformalgaugechoice} cannot be made because $\xi_2=2$ is not allowed, as discussed below \eqref{laggaugefixxie}.  Furthermore, the conformal descendent expression 
\be {\cal P}_{\mu_1\mu_2\mu_3\mu_4}^{ \rho_1\rho_2\rho_3\rho_4 } {\cal P}_{\nu_1\nu_2\nu_3\nu_4}^{ \sigma_1\sigma_2\sigma_3\sigma_4 } \partial_{\rho_1} \partial_{\rho_3} \partial_{\sigma_1} \partial_{\sigma_3} \left({1\over x^2}{\cal I}_{\rho_2\rho_4\ \sigma_2\sigma_4} \right), \ee
suggested by \eqref{WWcorrelatoree}, \eqref{hconf}, \eqref{Imunudefe} vanishes identically in $D=4$.

Another sign is that the unitarity bound for a primary with the symmetries of the Weyl operator is $\Delta\geq D-1$ \cite{Ferrara:2000nu,Costa:2014rya}. The dimension of $W$ is ${1\over 2}(D+2)$, which saturates this unitarity bound when $D=4$. At the unitarity bound, a conformal primary field with the symmetries of $W$ satisfies conservation conditions which are exactly of the form \eqref{conservqtioncondiee1}, \eqref{conservqtioncondiee2}. These conditions are thus conformally invariant in $D=4$, and so the linearized gravitational equations of motion, when expressed in terms of the gauge-invariant curvatures, are conformally invariant, as has been known in various formulations (including twistor formulations) \cite{Bracken:1982ny,Eastwood:1987ki,Siegel:1988gd,Erdmenger:1997gy,Shaynkman:2004vu,Nakayama:2013is,Barnich:2015tma,Flores:2017yyj,Flores:2017ubo}. Furthermore, insertion of the conservation equation into the correlation function commutes with the conformal Ward identities, which fix the conformal structure of two point correlators.  (The same observation can be made in the analogous situation of a spin one field, i.e. Maxwell theory.)

This leads us to suspect that $W$ is a conformal primary in $D=4$, and the theory is conformally invariant.
Computing the correlator explicitly in $D=4$ using any gauge choice other than \eqref{conformalgaugechoice}, we find\footnote{Putting the correlator in this form requires more than just tensor manipulations, it requires the use of dimensionally dependent identities for $D=4$. We accounted for this by evaluating everything explicitly in terms of components.}
\begin{align} 
\label{WWcorrelatoree4de} 
\la W_{\mu_1\mu_2\mu_3\mu_4}(x)& W^{\nu_1\nu_2\nu_3\nu_4}(0) \ra = \frac{96}{\pi^2 x^6} I^{\rho_1}_{\ \mu_1} I^{\rho_2}_{\ \mu_2} I^{\rho_3}_{\ \mu_3} I^{\rho_4}_{\ \mu_4} { \cal P}_{\rho_1\rho_2\rho_3\rho_4}^{\nu_1\nu_2\nu_3\nu_4},
\end{align}
where $I^\mu_{\ \nu}$ is as defined in \eqref{Imunudefe}.
This is precisely the conformally invariant form for a primary of weight $\Delta=3$ with the symmetries of the Weyl operator \cite{Osborn:1993cr,Costa:2014rya}. It satisfies the conservation conditions \eqref{conservqtioncondiee1}, \eqref{conservqtioncondiee2} expected of an operator saturating the unitarity bound $\Delta\geq D-1$. 

Since all other correlators are computed by Wick contraction from \eqref{WWcorrelatoree4de}, they will also automatically be conformally invariant. We thus conclude that linearized gravity, and hence the IR fixed point of quantum gravity, is in fact a CFT in $D=4$. It is a generalized free conformal field theory built from the Weyl primary. Like the $D>4$ case discussed above, it has no stress tensor, and no relevant or marginal scalar operators.

\section{Higher spin\label{HSsection}}

The conformal invariance we have discussed for the linearized Einstein-Hilbert action (which describes a free propagating massless spin-2 particle) mirrors what was found for Maxwell theory (which describes a free propagating spin-1 particle) in \cite{El-Showk:2011xbs}. While scale invariant in any $D$, Maxwell theory is conformal only in $D = 4$.  The difference is that Maxwell theory displays this conformal symmetry in the standard manner at the level of the action in $D=4$, whereas the linearized Einstein-Hilbert theory does not. For $D > 4$, it was shown in \cite{El-Showk:2011xbs} that Maxwell theory can be embedded into a larger conformally invariant theory by fixing a `conformal gauge,' in which the gauge field $A_\mu$ itself can be taken as a primary operator. Unitarity is lost because the conformal dimension of $A_\mu$, which is $1/2(D-2)$, lies below the unitary bound $\Delta\geq D-1$ for a vector primary, analogous to the gravitational case. In that work, the gauge-fixing ghosts were kept, which gives an even larger theory with a BRST symmetry whose singlet sector is the scale invariant Maxwell theory. We could presumably do the same for the linearized Einstein theory, though we can do without the ghosts or a BRST symmetry because a symmetry with a singlet sector, while useful, is not necessary for a consistent truncation to the scale and gauge-invariant theory \cite{Dymarsky:2015jia}. 

These statements should also generalize to the free theory of an arbitrary spin $s>2$ field.
A massless spin-$s$ particle can be described by a symmetric double-traceless gauge field $h_{\mu_1\cdots \mu_s}$ satisfying a gauge symmetry $\delta h_{\mu_1\cdots \mu_s}=\partial_{(\mu_1}\xi_{\mu_2\cdots \mu_s)}$ with a traceless symmetric rank $s-1$ gauge parameter $\xi_{\mu_1\cdots \mu_{s-1}}$ \cite{Fronsdal:1978rb}. The lowest dimension gauge-invariant operator is the generalized Weyl tensor field strength which involves $s$ derivatives of the gauge field \cite{deWit:1979sib},
\be W_{\mu_1\nu_1,\cdots ,\mu_s\nu_s }\propto \partial_{\mu_1}\cdots \partial_{\mu_s} h_{\nu_1\cdots\nu_s}+\cdots.\ee
This operator has the symmetries of a traceless Young tableaux with two rows of length $s$ and satisfies conservation conditions analogous to \eqref{conservqtioncondiee1} and \eqref{conservqtioncondiee2}. 

The generalized Weyl operator $W$ has dimension $\Delta = D/2+s-1$ and the unitarity bound on a conformal primary with its index symmetries is $\Delta \geq D+s-3$ \cite{Ferrara:2000nu,Costa:2014rya}. This inequality is saturated for $W$ only when $D=4$, and when $D=4$ the spin-$s$ equations of motion in their field strength formulation are conformally invariant. Thus in $D = 4$ the Weyl operator correlators of the free spin-$s$ theory should also be conformally invariant, and the theory should be a conformal field theory. In $D>4$ the free spin-$s$ action is scale but not conformally invariant, but we expect there should be a conformal choice of gauge-fixing which allows it to be embedded in a non-unitary conformal theory with a spin-$s$ primary of dimension $1/2(D-2)$ (violating the unitarity bound $\Delta\geq D+s-2$ of a spin-$s$ operator).  This conformal gauge-fixing should be unreachable when $D=4$.

\section{Conclusions}

We have seen that the linearized Einstein-Hilbert action \eqref{masslessefreeaction} is an example of a unitary scale but not conformally invariant theory in $D>4$. We have also shown explicitly that the correlation functions of the Weyl tensor operator derived from the action are those of a conformal primary in $D = 4$, even though the action is not invariant under conformal transformations in the standard manner (though it may be visible in non-local formulations \cite{Roiban:2017iqg}). To our knowledge, the fact that this action is consistent with a generalized free CFT in $D = 4$ was known implicitly in the literature but not demonstrated with an explicit derivation.  We also showed that this theory can be embedded into a non-unitary conformal theory for $D>4$ by judiciously gauge-fixing the action.

A central remaining question is whether conformal symmetry at the infrared fixed point can be at all useful when moving away from the fixed point, where $M_P$ turns on and breaks scale invariance. There are some indications that conformal symmetry, perhaps in a non-standard form, plays a role in fully non-linear gravity, for example in scattering amplitudes at tree level \cite{Loebbert:2018xce} and in soft theorems for gravity \cite{DiVecchia:2015jaq} in any $D$, but there are still many open questions about the full role of conformal symmetry in gravity.

\vspace{-.2cm}
\paragraph{Acknowledgments:} We would like to thank Glenn Barnich, James Bonifacio, Austin Joyce, Markus Luty, Yu Nakayama and Slava Rychkov for helpful conversations and comments.  KH acknowledges support from DOE grant DE-SC0009946, KH and KF acknowledge support from Simons Foundation Award Number 658908. OH was supported by the FWO-Vlaanderen through the project G006119N
and by the Vrije Universiteit Brussel through the Strategic Research Program “High-Energy Physics”.


\renewcommand{\em}{}
\bibliographystyle{utphys}
\addcontentsline{toc}{section}{References}
\bibliography{WeylConformal-arxiv-2}

\end{document}